\documentstyle[12pt,psfig,epsfig,aaspp4,amsfonts]{article}
\def\degr{\hbox{$^\circ$}}
\def\arcmin{\hbox{$^\prime$}}
\def\arcsec{\hbox{$^{\prime\prime}$}}

\def\fs{\hbox{$.\!\!^{\rm s}$}}

\def\lsim{\mathrel{\hbox{\rlap{\lower.55ex \hbox {$\sim$}}\kern-.0em
\raise.4ex \hbox{$<$}}}} 
\def\gsim{\mathrel{\hbox{\rlap{\lower.55ex \hbox {$\sim$}}\kern-.0em
\raise.4ex \hbox{$>$}}}} 
\newcommand{\gpm}[3]{$#1^{+#2}_{-#3}$}
\DeclareMathSymbol{\blacksquare}  {\mathord}{AMSa}{"04}
\DeclareMathSymbol{\blacktriangle}      {\mathord}{AMSa}{"4E}

\begin{document}

\title{The radio to X-ray spectrum of GRB 970508 on May 21.0 UT}
\author{T.J. Galama\altaffilmark{1}, R.A.M.J. Wijers\altaffilmark{2},
M. Bremer\altaffilmark{3}, P.J. Groot\altaffilmark{1},
R.G. Strom\altaffilmark{1,4}, C. Kouveliotou\altaffilmark{5,6}, J. van 
Paradijs\altaffilmark{1,7}} \altaffiltext{1}{Astronomical Institute
`Anton Pannekoek', University of Amsterdam, \& Center for High Energy
Astrophysics, Kruislaan 403, 1098 SJ Amsterdam, The Netherlands}
\altaffiltext{2}{Institute of Astronomy, Madingley Road, Cambridge,
UK} \altaffiltext{3}{Institut de Radio Astronomie Millim\'{e}trique,
300 rue de la Piscine, F--38406 Saint-Martin d'H\`{e}res, France}
\altaffiltext{4}{NFRA, Postbus 2, 7990 AA Dwingeloo, The Netherlands}
\altaffiltext{5}{Universities Space Research Asociation}
\altaffiltext{6}{NASA/MSFC, Code ES-84, Huntsville AL 35812, USA}
\altaffiltext{7}{Physics Department, University of Alabama in
Huntsville, Huntsville AL 35899, USA}

\begin{abstract}
We have reconstructed the spectrum of the afterglow of GRB 970508, on
May 21.0 UT (12.1 days after the GRB), on the basis of observations
spanning the X-ray to radio range.  The low-frequency power law index
of the spectrum, $\alpha=0.44\pm0.07$ ($F_\nu\propto\nu^\alpha$), is
in agreement with the expected value $\alpha=1/3$ for optically thin
synchrotron radiation. The 1.4 GHz emission is self-absorbed. 
We infer constraints on the
break frequecies $\nu_{\rm c}$ and $\nu_{\rm m}$ on May 21.0 UT from a
spectral transition from $F_{\nu} \sim \nu^{-0.6}$ to $F_{\nu} \sim
\nu^{-1.1}$ in the optical passband around 1.4 days.
A model of an adiabatically expanding blast wave,
emitting synchrotron radiation, in which a significant fraction of the
electrons cool rapidly provides a successful and consistent
description of the afterglow observations over nine decades in frequency,
ranging in time from trigger until
several months later.
\end{abstract}



\section{Introduction}
GRB970508 was a moderately bright
$\gamma$-ray burst (Costa et al. 1997; Kouveliotou et al. 1997). It was
detected on May 8.904 UT with the Gamma-Ray Burst Monitor (GRBM;
Frontera et al. 1991), and with the Wide Field Cameras (WFCs; Jager et
al. 1995) on board the Italian-Dutch X-ray observatory BeppoSAX (Piro
et al. 1995). Optical observations of the WFC error box (Heise et
al. 1997), made on May 9 and 10, revealed a variable object at RA =
$06^{\rm h}53^{\rm m}49\fs2$, Dec = +79\degr16\arcmin19\arcsec
(J2000), which showed an increase by $\sim$1 mag in the V band (Bond
1997), and whose spectrum indicated that the distance to the GRB source
corresponds to a redshift of at least 0.83 (Metzger et al. 1997).
The BeppoSAX Narrow Field Instruments (NFIs) revealed an X-ray
transient (Piro et al. 1997a) whose position is consistent with that
of the optical variable, and Frail et al. (1997a,b) found the first
GRB radio afterglow for GRB 970508; the radio source position
coincides with that of the optical source (Bond 1997).

The global properties of the X-ray, optical and radio afterglows of
GRBs have been successfully described by relativistic blast wave models
(e.g. Wijers, Rees and M\'esz\'aros 1997). In this paper we show that
the  detailed properties of the  
spectrum of the afterglow of GRB 970508, and its evolution, are well
described by a blast wave model in which significant cooling of the
electrons plays an important role (Sari, Piran and Narayan 1998). This
model predicts the occurrence of 
an extra break at $\nu_{\rm c}$ in the spectrum, corresponding to the
Lorentz factor $\gamma_{\rm c}$ above which the electron cooling time
is shorter than the expansion time of the blast wave.

\section{The X-ray to radio spectrum on May 21.0 UT}

In the second week after the GRB 970508 event many observations of
its afterglow were made by ground-based and orbiting instruments, which allows
us to derive the X-ray to radio 
afterglow spectrum on May 21.0 UT (12.1 days after the event)  with
only a few assumptions.   

During the first month the 4.86 and 8.46 GHz observations by Frail et
al. (1997a) show strong fluctuations attributed to interstellar
scintillation (ISS; see e.g. Goodman 1997).  At 8.46 GHz the interval
of 8 to 14 days after the event corresponds to a flare. It would
therefore appear that the `true' underlying 4.86 and 8.46 GHz flux
densities are better represented by long-term average values. To
discern, apart from an average flux, a possible trend in the 8.46 and
4.86 GHz data of Frail et al. (1997a) we have fitted a power law (in
time $t$, measured in days since the burst) to the data for $t <$ 30
days. We find $F_{\rm 4.86GHz} = (101 \pm 17)\times t^{(0.61\pm0.06)}$
$\mu$Jy ($\chi_{\rm red}^{2} = 254/10)$ and $F_{\rm 8.46GHz} = (321
\pm 18)\times t^{(0.274\pm0.020)}$ $\mu$Jy ($\chi_{\rm red}^{2} =
521/25)$ and we have used these expressions to derive a flux estimate
for May 21.0 UT at these frequencies. At lower frequencies the
afterglow was not (or barely) detected during this week. The average
Westerbork Synthesis Radio Telesecop (WSRT) 1.4 GHz flux density (May
9.7-Jul 16.44 UT) is $<$ 80 $\mu$Jy (2$\sigma$; Galama et al. 1998a;
we note, however, that Frail et al. 1997a report a flux density of
$\sim$ 100 $\mu$Jy).

Pooley and Green (1997) observed GRB 970508 on six consecutive days
(May 17-22) with the Ryle telescope at 15 GHz. The source was detected
on May 16.52 UT at 1.57 $\pm$ 0.25 mJy, but on none of the other days
was an individually significant detection made. However, averaging all
observations obtained in the interval May 17-22 (8-13 days after the
event), the source was detected at 660 $\pm$ 110 $\mu$Jy. Again, since
ISS can modulate the flux densities strongly below$\sim20$~GHz at this
galactic latitude, we assume that the detection on May~16.52 UT was
boosted by this effect, and that the source-intrinsic 15~GHz flux on
May~21.0 is better represented by the five-day average stated above.

Bremer et al. (1998) detected GRB 970508 with the IRAM Plateau de Bure
Interferometer (PdBI) at 86 GHz but not at 232 GHz.  No significant
variation is observed between the three 86 GHz PdBI detections (May
19.07-May 22.99; center epoch May 21.0 UT) and, combining them we
obtain an average $F_{\rm 86 GHz}$ = 1.71 $\pm$ 0.26 mJy.  For the 232
GHz upper limit we use the May 21.15 (day 12.2) upper limit of $F_{\rm
232 GHz} < 4.22$ mJy (2$\sigma$; Bremer et al. 1998).

In an observation with the Infrared Space Observatory (ISO)
on May 21 the source was not detected (Hanlon et al. 1998):
preliminary upper limit  $F_{12\mu{\rm m}} < 120 \mu$Jy.

We have fitted a power law, $F_{\rm K}$ = \gpm{120}{75}{48} $\times t^{-0.89
\pm 0.27} \mu$Jy ($\chi_{\rm red}^{2} = 0.23/1)$, to the three K band
(2.2 $\mu$m) detections of Chary et al. (1998), and find $F_{\rm K}$ =
\gpm{13}{15}{8} $\mu$Jy for May 21.0 UT.

For an estimate of the optical flux density we used the fit by Galama
et al. (1998b) to the power-law decay of the differential Coussins R 
(R$_{\rm c}$;  $\lambda_{\rm c} $ = 6400 \AA)  light
curve, i.e., $F_{\rm
R_{\rm c}}= (82.7 \pm 2.3) \times t^{-1.141 \pm 0.014}$ $\mu$Jy; using
an absolute calibration uncertainty of 0.1 mag we find 4.8 $\pm$ 0.5
$\mu$Jy.

X-ray fluxes (2-10 keV) of the afterglow, obtained with BeppoSAX, have
been reported by Piro et al. (1997b). These observations were made
during the burst, and between 6 hours and 6 days afterwards. We
have inferred the X-ray flux $F_{\rm X}$ = (7.3 $\pm$ 4)$\times
10^{-4} \mu$Jy for May 21.0 UT by extrapolating the power law fit
($F_X \propto t^{-1.1 \pm 0.1}$) of Piro et al. (1997b).

\section{Discussion \label{Disc}}

\subsection{The radio to X-ray spectrum}
\label{RXspec}
The radio to X-ray spectrum on May 21.0 UT is shown in
Fig. \ref{fig:spec}. We have divided the spectrum into four parts,
corresponding to: the lowest frequencies (I; $< 2.5 \times 10^{9}$
Hz), the low frequencies (II; 2.5$\times 10^{9}$ - $10^{11}$ Hz), the
intermediate frequencies (III; $10^{11}$ - $10^{14}$ Hz) and the high
frequencies (IV; $>$ $10^{14}$ Hz).  Region I has a much steeper
spectral slope, $\alpha_{\rm 1.4-4.86 GHz} > $ 1.1, than region II,
$\alpha_{\rm 4.86-86 GHz}$ = 0.44 $\pm$ 0.07 ($\chi^2_{\rm red}$ =
1.4/2).  The 1.4 GHz flux density (region I) is consistent with the
expectation for self-absorbed emission ($F_{\nu} \propto \nu^2$; Katz
and Piran 1997), i.e. the self absorption break $\nu_{\rm a}$ $\sim$
2.5 GHz. The low-frequency region (II) is in agreement with the
expected low-frequency tail of synchroton radiation ($F_{\nu} \propto
\nu^{1/3}$; Rybicki and Lightman 1979).  The high frequency X-ray to
optical slope $(\alpha = -1.12 \pm 0.07)$ is consistent (Galama et
al. 1998a; from here on Paper I) with synchroton radiation from
electrons, of which a significant fraction cools rapidly, with a
power-law distribution of Lorentz factors, $N(\gamma_{\rm e}) \propto
\gamma_{\rm e}^{-p}$ (Sari et al. 1998).  Optical multi-colour
photometry, obtained in the first 5 days after the event, showed that
the optical spectrum was well represented by a power law, and
approached $\alpha$ = $-1.11 \pm 0.06$ (2.1 to 5.0 days after the GRB;
$\chi_{\rm red}^{2} = 3.0/4)$ after reaching maximum optical light
(Galama et al. 1998b; we used $t >$ 2.1 days), i.e., consistent with
the X-ray to optical slope on May 21.0 UT, suggesting that the slope
remained constant over this period.  We note that the extrapolation of
the X-ray flux is uncertain, but none of the further discussion
depends on the validity of the extrapolation, since both spectral
breaks we infer lie below the optical.  In the spectrum the local
optical spectral slope derived from 
Galama et al. (1998b) is also indicated.

For region III, only upper limits are available. Extrapolating from
the adjacent regions and assuming one single spectral break, we find
a peak flux of 4.5~mJy at $10^{12}$Hz. However, the temporal
development of GRB 970508 indicates that a second break frequency exists.

\subsection{Evidence for a second spectral break}
\label{Opt}
As noted before by Katz, Piran and Sari (1998), during the first day
the optical to X-ray 
spectral slope of GRB970508 $\alpha \sim -0.5$; which
these authors attribute to rapid electron cooling. This
is consistent with the UV excess reported by Castro-Tirado et al. (1998),
and the reddening of the optical spectrum during the first five days (Galama
et al. 1998b).  
From the data of Galama et al. (1998b) we determine (using $A_V <0.01$,
based on the 
IRAS 100$\mu$m cirus flux, instead of the value $A_V = 0.08$ used by
Galama et al. 1998b) $\alpha = -0.54 \pm 0.14$ for $t$ between 0 and 1.5
days, and $\alpha = -1.12 \pm0.04$ for $t$ between 1.5 and 5 days. 
This 
suggests that a spectral break moved through the optical passband
between 1.0 and 1.8 days after the burst, separating a range with
slope $\alpha \sim -0.5$ from one with slope $\alpha \sim -1.1$ (see
Fig. \ref{fig:trans} upper panel). 
By May 21 we expect this break to have moved downward in frequency to
$\sim 10^{14}$ Hz (see Fig. 
1 and Sect. \ref{sec:cons}).
Additional observational evidence for rapid cooling of a significant
fraction of the	 
electrons is given in Paper I.

\subsection{Rapid electron cooling}
Sari et al. (1998)  argued that the most energetic electrons may lose a
significant fraction of their energy to radiation.  
Their model predicts the occurrence of  an extra break at
$\nu_{\rm c}$ (corresponding to the critical Lorentz factor
$\gamma_{\rm c}$ above which cooling by synchroton radiation
is significant), in addition to the regular break
frequency $\nu_{\rm m}$ (the frequency which corresponds to the
lowest-energy injected electrons; Lorentz factor $\gamma_{\rm m}$).
The evolution in time of the GRB afterglow is determined by the
evolution of the two break frequencies: $\nu_{\rm c} \propto t^{-1/2}$
and $\nu_{\rm m} \propto t^{-3/2}$ (Sari et al. 1998). We assume
adiabatic evolution of the GRB remnant; see Sect. \ref{adiab}.  Since
$\nu_{\rm m}$ is initially greater than $\nu_{\rm c}$ (Sari et
al. 1998), but decays more
rapidly, there is a time, $t_0$, when the two are equal. For late
times $t > t_0$ we have $\nu_{\rm m} < \nu_{\rm c}$ and
the spectrum varies as $F_{\nu} \propto \nu^{-(p-1)/2}$ from $\nu_{\rm
m}$ up to $\nu_{\rm c}$; above $\nu_{\rm c}$ it follows $F_{\nu}
\propto \nu^{-p/2}$ and below $\nu_{\rm m}$ it follows the 
low-frequency tail, $F_{\nu} \propto \nu^{1/3}$ (Sari et al. 1998).

\subsection{Constraints on the break frequencies $\nu_{\rm c}$ and
$\nu_{\rm m}$ for May 21.0 UT} 
\label{sec:cons}
The 86 GHz PdBI observations do not detect the source before 10 days,
followed by three detections and subsequently only non detections
i.e. at 86 GHz the emission peaked between $\sim$ 10 and $\sim$ 14
days (Bremer et al. 1998).  This implies that the peak flux
$F_{\nu,{\rm max}} \sim 1700 \mu$Jy at 86 GHz.  The decay after the
maximum is quite rapid: for $F_{\nu} \propto t^{\delta}$ we find
$\delta < $ --1.1 using the  3$\sigma$ upper limit on May 28.41
(Bremer et al. 1998).  If this maximum would correspond to the break
frequency $\nu_{\rm c}$ 
passing 86 GHz then we would expect the decay to be much slower, $F_{\nu}
\propto t^{-1/4}$ (Sari et al. 1998). If it reflects the passage of
$\nu_{\rm m}$ the subsequent decay would go as $F_{\nu}
\propto t^{3(1-p)/4} = t^{-0.9}$ (Sari et al. 1998). The observed
decay rate is in
marginal agreement with the passage of the break frequency $\nu_{\rm
m}$, but definitely excludes 
that $\nu_{\rm c}$ was passing 86 GHz. We therefore identify the
maximum at 86 GHz with the passage of $\nu_{\rm m}$ at $t_{{\rm m,86
GHz}}$ $\sim$ 12 days, 
i.e. at the time of the
derived spectrum (May 21.0 UT).

\begin{figure}[ht]
\centerline{\psfig{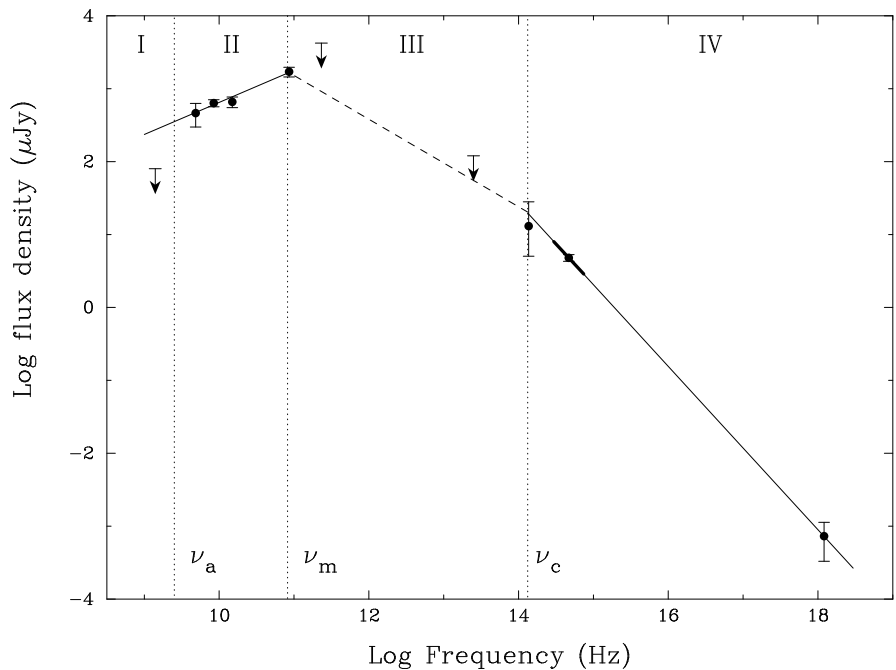}}
\caption[]{The X-ray to radio spectrum of GRB 970508 on May 21.0 UT (12.1
days after the event). The fit to the low-frequency part,
$\alpha_{\rm 4.86-86 GHz}$ = 0.44
$\pm$ 0.07, is shown as well as the extrapolation from X-ray to
optical (solid lines). The local optical slope (2.1--5.0 days after
the event) is indicated by the thick solid line. Also indicated is
the extrapolation $F_{\nu} \propto    
\nu^{-0.6}$ (lines). Indicated are the rough estimates of the break
frequencies $\nu_{\rm a}$, $\nu_{\rm m}$ and $\nu_{\rm c}$ for May 21.0 UT. 
\label{fig:spec}}  
\end{figure}

There is no evidence for a
break at $\nu < 86$ GHz (the 86 GHz detection is consistent with the
low-frequency tail; see Fig. \ref{fig:spec}). This implies that
$\nu_{\rm c} > \nu_{\rm m}$ and 
so for $\nu > \nu_{\rm m}$ the spectrum is predicted to follow $F_{\nu}
\propto \nu^{-(p-1)/2} = \nu^{-0.6}$ (we will use $p$ = 2.2, Paper I).
Note also that therefore $t > t_0$ at 12.1 days. 
The intersection of the high frequency extrapolation with $F_{\nu}
\propto \nu^{-0.6}$ for $\nu > \nu_{\rm m}$ gives an estimate for
$\nu_{\rm c}$ (see Fig. \ref{fig:spec}); we find $\nu_{\rm c}$ =
$1.6\times10^{14}$ Hz at 12.1 days (May 21.0 UT).  Independently, we
identify the observed optical spectral transition between 1.0 and
1.8 days from $\alpha$ = $-0.54 \pm 0.14$ to $\alpha$ = $-1.12 \pm
0.04$ (Sect. \ref{Opt} and Fig. \ref{fig:trans}) with the break
frequency $\nu_{\rm c}$ passing 
the R$_{\rm c}$ band at $t_{{\rm
c,R_c}}$. For $t < t_{{\rm c,R_c}}$ we expect $F_{\nu} \propto
\nu^{-0.6}$, while for $t > t_{{\rm c,R_c}}$ we expect $F_{\nu} \propto
\nu^{-p/2}=\nu^{-1.1}$ in the optical passband; in excellent agreement
with the observed transition. Using $t_{{\rm c,R_c}}$ = 1.4 days
($\nu_{\rm c} \propto t^{-1/2}$), we find $\nu_{\rm c}$ = 1.6 $\times
10^{14}$ Hz at $t$ = 12.1 days (May 21.0 UT). This is in remarkable
agreement with the value derived above from the intersection of
$F_{\nu} \propto \nu^{-0.6}$ for $\nu > \nu_{\rm m}$ in region III 
with the high-frequency extrapolation. We therefore identify the
transition observed 
in the optical passband with the passage of the 
break frequency $\nu_{\rm c}$ at $t_{{\rm c,R_c}} \sim$ 1.4 days.  The
frequencies   
$\nu_{\rm a}, \nu_{\rm m}$ and $\nu_{\rm c}$ are shown in 
Fig. (\ref{fig:spec}); they determine the division of the spectrum
into the four regions.  We note that the upper limit given for the ISO
non detection (Hanlon et al. 1998) is a conservative number and
further analysis may provide a stronger constraint on the shape of the
spectrum.

\subsection{The near-infrared range}
\label{sec:infra}
\begin{figure}[ht]
\centerline{\psfig{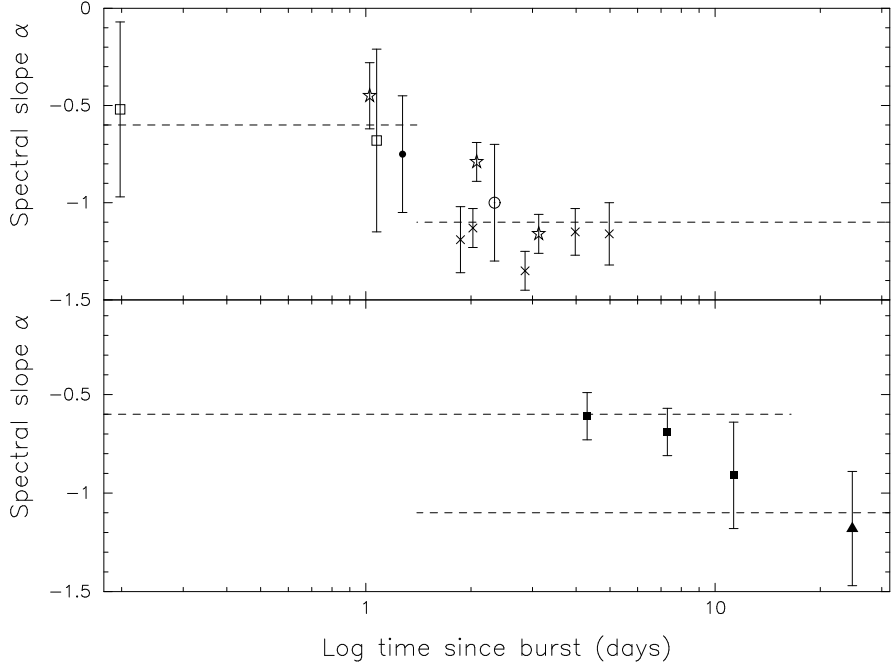}}
\caption[]{The spectral slope $\alpha$ as a function of the logarithm
of time in days after the burst. The upper panel is from Galama et
al. (1998b) and shows the spectral
slope transition in the optical passband (slopes determined from U, B,
V, R$_{\rm c}$ and I$_{\rm c}$ band observations; we used $A_V$ = 0, see
Sect. \ref{Opt}), where data is collected from Galama et
al. (1998b; $\star$), Sokolov et al. (1997; $\times$), Djorgovski et
al. (1997, $\bullet$), Metzger et al. (1997b, $\circ$) and
Castro-Tirado et al (1998; $\Box$).  The transition from $F_{\nu}
\propto \nu^{-0.6}$ to $F_{\nu} \propto \nu^{-1.1}$ (dashed lines)
occurs at roughly 1.4 days.  We expect a
similar transition in the spectral slope derived from K-R$_{\rm c}$
(lower panel) to occur between 1.4  
and 16.5 days (see Sect. \ref{sec:infra}). The
lower panel shows the spectral indices $\alpha_{\rm K-R_{\rm c}}$
($\blacksquare$), derived from the K band data of Chary et al. (1998), and
$\alpha_{\rm H-R_c}$  ($\blacktriangle$), derived from the H band data
of Pian et 
al. (1998), with respect to the the optical R$_{\rm c}$ light curve of
Galama et al. (1998b).}
\label{fig:trans}
\end{figure}

At 2.2$\mu$m (K band) we expect $\nu_{\rm c}$ to pass at $t_{\rm c,K}
\sim$ 16.5 days ($\nu_{\rm c} \propto t^{-1/2}$). Before that the K
band flux density should decay as $F_{\nu} \propto t^{-0.9}$ (see
above). The decay $F_{\rm K} \propto t^{-0.89 \pm 0.27}$
(Sect. \ref{RXspec}), fitted to the data of Chary et al. (1998), is in
good agreement with this. The K to R band spectrum should show a
gradual transition between 1.4 days (when $\nu_{\rm c}$ passes R$_{\rm
c}$) and 16.5 days from $F_{\nu} \propto \nu^{-0.6}$ to $F_{\nu}
\propto \nu^{-1.1}$. Spectral indices $\alpha_{\rm K-R_{\rm c}}$,
derived from the K band data of Chary et al. (1998) and the optical
$R_{\rm c}$ light curve of Galama et al. (1998b) are in good agreement
with this: $\alpha_{\rm K-R_{\rm c}}$ = $-0.61 \pm 0.12$ ($t=$ 4.3),
$\alpha_{\rm K-R_{\rm c}}$ = $-0.69 \pm 0.12$ ($t=$ 7.3), and
$\alpha_{\rm K-R_{\rm c}}$ = $-0.91 \pm 0.27$ ($t=$ 11.3; see
Fig. \ref{fig:trans} lower panel). H band (1.6 $\mu$m) HST
observations (Pian et al. 1998) detect the OT at H = 20.6 $\pm$ 0.3
($F_{\rm H} = 6.2 \pm 1.5 \mu$Jy, $t=$ 24.7) and comparison with
Galama et al. (1998b) gives $\alpha_{\rm H-R_c} = -1.18 \pm 0.29$
(Fig. \ref{fig:trans} lower panel).  This is consistent with the
expectation that $\nu_{\rm c}$ had by then already passed the H band.

\subsection{Adiabatic dynamical evolution of the blast wave and the value of $t_0$}
\label{adiab}
The observed transitions $t_{\rm m,86 GHz}$ $\sim$ 12 days and
$t_{\rm c,R_c} \sim$ 1.4 days imply that $t_0 \sim$ 0.006 days
(500 s).  For $t>t_0$ the dynamical evolution of the remnant is
almost certainly adiabatic, while before that time it could possibly
be radiative (Sari et al. 1998; but see M\'esz\'aros, Rees \& Wijers
1998).  This means that one would expect nearly all of the remnant's
observed dynamical evolution to be adiabatic.

However, the maximum flux at 8.46 GHz is
$F_{\nu,{\rm max}} \sim 700 \mu$Jy at $t_{\rm m,8.46 GHz} \sim$ 55 days (Frail
et al. 1997a); this value of $t_{\rm m,8.46 GHz}$ is as expected from
the 86 GHz observations, for $\nu_{\rm m} \propto t^{-3/2}$.
This is less than the peak of 
1700 $\mu$Jy at 86 GHz and would argue for some radiative losses,
since in the perfectly adiabatic case the maximum flux should be constant
with time, while the peak moves to lower frequencies. 
Also the rather gradual early evolution of the radio 
light curves of GRB 970508 and the observed transition from optically thick to
thin emission at 1.4 GHz around $t \sim 45$ days suggest the
possibility of radiative evolution (Paper I). 
However, the absence of a break in the smooth
power law decay of the optical light curve from 2 to 60 days after the
burst (Pedersen et al. 1998; Castro-Tirado et al. 1998; Sokolov et
al. 1998; Galama et al. 1998b) shows that there is no important
transition in that period. Also the optical spectra and the optical
temporal slope
indicate adiabatic remnant dynamics with $\nu_{\rm c}>\nu_{\rm m}$
during this period (Paper I). Taken together, these results indicate
that some additional ingredient is needed to fully explain the radio
and mm behaviour; for example,
Waxman, Kulkarni and Frail (1998) have argued that the transition from
ultrarelativistic to mildly relativistic expansion of the blast wave
may explain these anomalies.

\section{Conclusion}

We have found that on May 21.0 UT the afterglow spectrum of
GRB 970508 contains all the characteristic parts of a synchrotron
spectrum. We infer that
the dynamical evolution of the remnant is adiabatic (in agreement with
Waxman et al. 1998), but that a significant fraction of the electrons
have synchrotron cooling 
times shorter than the remnant's expansion time. In accordance with the theory 
outlined by Sari et al. (1998)  this produces an additional 
break in the spectrum. The break frequencies $\nu_{\rm m}$ and $\nu_{\rm c}$,
corresponding to the minimum electron energy and the
energy above which 
electrons cool rapidly, respectively, are equal at a time $t_0$ which
we find is 500 s. This synchrotron
spectrum and adiabatic dynamics
can explain most of the afterglow behaviour over 9 decades in
frequency spanning the X rays to radio waves, from the 
first optical to the late radio data. Some aspects of the afterglow are
not explained by this model: the flux at $\nu_{\rm m}$ 
should be constant, but is seen to decrease from 12 days when it lies at
mm wavelengths to 55 days when it lies in the radio; the self-absorption
frequency $\nu_{\rm a}$ is predicted to be time-independent, but is
seen to decrease 
with time and the early rise of the radio fluxes is slower than
expected (see Paper I) and finally the decay after maximum at mm
wavelengths is perhaps somewhat faster than expected (see paper I for
a discussion on the radio light curves).

In summary, we find that an adiabatically expanding blast wave emiting
synchrotron radiation, with a significant fraction of rapidly 
cooling electrons, describes most of the afterglow data on GRB 970508
very well.

\acknowledgments

We would like to thank Dr. Katz for useful discussions.
T. Galama is supported through a
grant from NFRA under contract 781.76.011.  R. Wijers is supported by
a Royal Society URF grant.  C. Kouveliotou acknowledges support from
NASA grant NAG 5-2560.

\newpage


\begin{references}



\reference{}Bond, H. 1997, \iaucirc, 6654
\reference{}Bremer, M., Krichbaum, T.P., Galama, T.J., Castro-Tirado,
A.J., Frontera, F., van Paradijs, J., Mirabel, I.F., Costa,
E. 1998, \aap, 332, L13
\reference{}Castro-Tirado, A.J. et al. 1998, Science, 279, 1011
\reference{}Chary, R. et al. 1998, accepted for publication in, \apjl
\reference{}Costa, E. et al. 1997, \iaucirc, 6649
\reference{}Frail, D.A., Kulkarni, S.R., Nicastro, L., Feroci, M.,
Taylor, G.B. 1997a, \nat, 389, 261 
\reference{}Frail, D.A. et al. 1997b, \iaucirc, 6662
\reference{}Frontera, F. et al. 1991, Adv. Space Res. 11, 281
\reference{}Galama, T.J. et al. 1998a, accepted for publication in,
\apjl (Paper I)
\reference{}Galama, T.J. et al. 1998b, \apjl, 497, 13
\reference{}Goodman, J. 1997, New Astronomy, 2, 449
\reference{}Hanlon, L. et al. 1998, in preparation (private communication)
\reference{}Heise. J. et al. 1997, \iaucirc, 6654
\reference{}In 't Zand, J. et al. 1997, \iaucirc, 6569
\reference{}Jager, R., Heise, J., In 't Zand, J., Brinkman, A.C.
1995, Adv. Space Res., 13, 315
\reference{}Katz, J.I. and Piran, T. 1997, \apj, 490, 772
\reference{}Katz, J.I., Piran, T. and Sari, R. 1998, \prl, 80, 1580
\reference{}M\'esz\'aros, P.I., Rees, M.J., and Wijers, R.A.M.J., 1997,
\apj, in press.
\reference{}Metzger, M.R., Djorgovski, S.G., Kulkarni, S.R., Steidel,
C.C., Adelberger, K.L., Frail, D.A., Costa, E., Frontera, F.  1997,
\nat, 387, 878  
\reference{}Pedersen, H. et al. 1998, \apj, 496, 311
\reference{}Pian, E. et al. 1998, \apjl, 492, 103
\reference{}Piro, L., Scarsi, L. and Butler, R.C. 1995, Proc. SPIE
2517, 169
\reference{}Piro, L. et al. 1997a, \iaucirc, 6656
\reference{}Piro, L. et al. 1997b, \aap, 331, L41
\reference{}Pooley, G. and Green, D. 1997, \iaucirc, 6670
\reference{}Rybicki, G.B., Lightman, A.P. 1979, Radiative Processes In
Astrophysics, Wiley, New York
\reference{}Sari, R., Piran, T. and Narayan, R. 1998, \apjl, 497, 17
\reference{}Sokolov, V.V., Kopylov, A.I., Zharikov, S.V., Feroci, M.,
Nicastro, L., Palazzi, E. 1998, accepted for publication, \aap
astro-ph 9802341
\reference{}Van Paradijs, J. et al. 1997, \nat, 386, 686
\reference{}Waxman, E., Kulkarni, S.R. \& Frail. D.A. 1998, \apj, 497, 288
\reference{}Wijers, R.A.M.J, Rees, M.J., and M\'esz\'aros, P.I. 1997,
\mnras, 288, L51

\end{references}
\end{document}